\documentclass[10pt,preprint]{article}
\usepackage{amsfonts,amssymb}

\newcommand{\bq}{\begin{equation}}
\newcommand{\nq}{\end{equation}}

\begin{document}

\title{Non-Markovian stochastic Liouville equation and anomalous relaxation kinetics. }

\author{A. I. Shushin}

\maketitle

\begin{center}
Institute of Chemical Physics, Russian Academy of Sciences, 117977, Kosygin str. 4, Moscow, Russia
\end{center}

\begin{abstract}
The kinetics of phase and population relaxation in quantum systems induced by noise with anomalously slowly decaying  correlation function $P (t) \sim (wt)^{-\alpha}$, where $0 <\alpha < 1$ is analyzed within continuous time random walk approach (CTRWA). The relaxation kinetics is shown to be anomalously slow. Moreover for $\alpha < 1$ in the limit of short characteristic time of fluctuations $w^{-1}$ the kinetics is independent of $w$. As $\alpha \to 1$ the relaxation regime changes from the static limit to fluctuation narrowing. Simple analytical expressions are obtained  describing the specific features of the kinetics.
\end{abstract}


\bigskip

{\bf Introduction.}\,
The noise induced relaxation in quantum systems is very important process observed in: magnetic resonance \cite{Abr}, quantum optics and nonlinear spectroscopy \cite{Muk}, etc.

Often these processes are analyzed assuming conventional stochastic properties of the noise: fast decay of correlation functions and short correlation time $\tau_c$ \cite{Abr}. In the absence of memory the relaxation is described by very popular Bloch-type equations.  As for memory effects, they are also discussed (within the Zwanzig projection operator approach \cite{Fors}), however, in the lowest orders in the fluctuating interaction $V$ inducing relaxation \cite{Argy}.

Recently strong attention has been drawn to the processes governed by noises with anomalously slowly decaying correlation functions $P (t) \sim t^{-\alpha}$ with $\alpha < 1$. They are discussed in relation to spectroscopic studies of quantum dots (\cite{Shim,Silb} and references therein). Similar problems are analyzed in the theory of stochastic resonances \cite{Han}.

Such anomalous processes cannot be properly described by methods based on expansion in powers of $V$. The goal of this work is to analyze the corresponding anomalous relaxation within the continuous time random walk approach (CTRWA) \cite{Hau} with the use of the recently derived non-Markovian stochastic Liouville equation (SLE) \cite{Shu1} which enables one to describe relaxation kinetics without expansion in $V$. In some physically reasonable models it allows for describing the phase and population relaxation kinetics in analytical form even for multilevel systems. In particular, the kinetics is shown to be strongly non-exponential .

\bigskip

{\bf General formulation.}
We consider noise induced relaxation in the quantum system whose dynamical evolution is governed by the hamiltonian
\bq
H (t) =  H_s + V (t),
\label{gen1}
\nq
where $ H_s$ is the term independent of time and $ V (t)$ is the fluctuating interaction, which models  effects of the noise. The evolution is described by the density matrix $\rho (t)$ satisfying the Liouville equation ($\hbar = 1$)
\bq
\dot \rho = - i {\hat H} (t) \rho, \;\: \mbox{with} \;\: {\hat H} \rho = [ H,\rho ] = [ H \rho - \rho H ].
\label{gen2}
\nq

${V} (t)$-fluctuations are assumed to be symmetric ($\langle V \rangle = 0$) and result from stochastic jumps between the states $|x_{\nu} \rangle$ in the (discrete or continuum) space $ \{ x_{\nu} \} \equiv \{ x \}$ with different $V = V_{\nu}$ and $H = H_{\nu}$ (i.e. different $\hat V = \hat V_{\nu} \equiv [V_{\nu}, \dots] $ and $\hat H = \hat H_{\nu}$):
\bq
\hat {\cal V}  = \sum\nolimits_{\nu} \! |x_{\nu} \rangle \hat V_{\nu} \langle x_{\nu} |
\;\,\mbox{and} \;\, \hat {\cal H}  = \sum\nolimits_{\nu} \! |x_{\nu} \rangle {\hat H}_{\nu} \langle x_{\nu} |.
\label{gen4}
\nq

Hereafter we will apply the bra-ket notation: $|k\rangle\,$ and  $|k k'\rangle \equiv |k\rangle \langle k' |$ for the eigenstates of $H$ (in the original space) and ${\hat H}$ (in the Liouville space), respectively, as well as notation $ |x\rangle$ for states in $ \{ x \}$-space.

The macroscopic evolution of the system under study is determined by the evolution operator ${\hat {\cal R}} (t)$ in the Liouville space averaged over $V (t)$-fluctuations:
\bq
\rho (t) = {\hat {\cal R}} (t) \rho_i \; \mbox{with} \;{\hat {\cal R}} (t)
= \!\sum\nolimits_{x, x_i} \!\!\hat {\cal G} (x,x_i|t) P_e (x_i),
\label{gen5a}
\nq
where $\hat {\cal G} (x,x'|t)$ is the average evolution operator and  $P_e (x)$ is the equilibrium distribution in $\{x\}$-space.

Non-Markovian $V (t)$-fluctuations will be described by the CTRWA [which leads to the non-Markovian SLE \cite{Shu1} for $\hat {\cal G} (t)$]. It treats fluctuations as a sequence of sudden changes of $\hat V $. The onset of any particular change of number $j$ is described by the matrix $\hat P_{j-1}$ (in $\{x\}$-space) of probabilities  not to have any change during time $t$ and its derivative $\hat W_{j-1} (t) = - d\hat P_{j-1} (t)/ dt$.  These matrices are diagonal and independent of $j$: $\hat P_{j-1} (t) = \hat P (t), \: \hat W_{j-1} (t) = \hat W (t) = - d\hat P (t)/ dt, \; (j > 1)$ , except $\hat P_{0} (t) \equiv \hat P_i (t)$ and $\hat W_{0} (t) \equiv \hat W_{i} (t) = - d\hat P_{i} (t)/ dt$ depending on the problem considered. For non-stationary ($n$) and stationary ($s$) fluctuations \cite{Hau} $\hat W_i (t) = \hat W_n (t) = \hat W (t)$ and $\hat W_i (t) = \hat W_s (t) = {\hat t_w}^{-1} \int_t^{\infty} \! d\tau \, \hat W (\tau)$, respectively, where $\hat t_w = \int_0^{\infty} \! d\tau \, \tau\hat W (\tau)$ is the matrix of average times of waiting for the change \cite{Hau}.

In what follows we will mainly operate with the Laplace transforms denoted as ${\widetilde Z}(\epsilon) = \int_0^{\infty} \! dt \, Z(t)e^{-\epsilon t}$ for any function $Z(t)$. In particular, noteworthy is the relation $\hat {\widetilde P}_j (\epsilon)= [1 - \hat {\widetilde W}_j (\epsilon)]/\epsilon$ and suitable representations
\bq
\hat {\widetilde {W}} (\epsilon) = [1 + \hat \Phi (\epsilon)]^{-1} \;\: \mbox{and} \;\: \hat {\widetilde P} (\epsilon) = [\epsilon + \epsilon/\hat \Phi (\epsilon)]^{-1}
\label{gen6}
\nq
in terms of a diagonal matrix $\hat \Phi (\epsilon)$ with $\hat \Phi (\epsilon) \stackrel{\epsilon \to 0}{\approx} (\epsilon/\hat w)^{\alpha}$, where $\hat w$ is a constant matrix and $\alpha \leq 1$ (see below).

Evolution in $ \{ x \}$-space is governed by the jump operator $\hat {\cal L} = 1- \hat {\cal P}$ in which $\hat {\cal P}$ is the the non-diagonal matrix of jump probabilities. Evolution leads to the equilibrium state $|e_x \rangle$, satisfying eq. $\hat {\cal L}\hat w^{\alpha} |e_x \rangle  = 0$ and represented as $|e_x \rangle = \sum_x P_e (x)|x \rangle$ with $\langle e_x | = \sum_x \langle x |$ \cite{Shu1}. Note that [see eq. (\ref{gen5a})]
\bq
{\hat {\cal R}} (t) = \langle e_x| \hat {\cal G} |e_x\rangle \equiv \langle \hat {\cal G} \rangle.
\label{gen6a}
\nq

The CTRWA leads to the non-Markovian SLE for $\hat {\cal G} (x, x_i|t)$ \cite{Shu1}. Solution of this SLE yields \cite{Shu1}
\bq
\hat {\widetilde {\cal G}} = \hat {\widetilde P}_i (\hat \Omega) + \hat \Omega^{-1} \hat
{\Phi} (\hat \Omega)  [\hat \Phi (\hat \Omega) + {\hat {\cal L}}]^{-1}
{\hat {\cal P}}\hat {\widetilde W}_{i} (\hat \Omega),
\label{gen7}
\nq
where
\bq
{\hat {\cal L}} = 1 - {\hat {\cal P}} \;\; \mbox{and} \;\; \hat \Omega = \epsilon + i \hat {\cal H}.
\label{gen8}
\nq
In particular, in the case of $n$-fluctuations ($\hat W_i = \hat W$)
\bq
\hat {\widetilde {\cal G}} = \hat {\widetilde {\cal G}}_n =
\hat \Omega^{-1} \hat \Phi(\hat \Phi + \hat {\cal L})^{-1}. \label{gen9}%
\nq
For $s$-fluctuations ($\hat W_i = \hat W_s $) $\hat {\widetilde {\cal G}} = \hat \Omega^{-1} - \hat {\widetilde {\cal G}}_n {\hat {\cal L}}(\hat \Omega \hat t_w)^{-1}$.

Hereafter, for brevity, we omit the argument $\hat \Omega$ of all Laplace transforms if this does not result in confusions.

\bigskip

{\bf Useful models and approaches.\,}

{\it Sudden relaxation model (SRM).}\,
The SRM \cite{Shu1} assumes sudden equilibration in $\{x\}$-space described by operator
\bq
\hat {\cal L} = (1 - |e_0 \rangle \langle e_0 |)\hat Q^{-1},\;\; \hat Q = 1-\mbox{$\sum\nolimits_x$}  P_0 (x) |x \rangle \langle x |,
\label{sud1}
\nq
in which $|e_0 \rangle = \sum_x\!  P_0 (x) |x \rangle,\: \langle e_0 |=\sum_x \langle x |.$ For this $\hat {\cal L}$
\bq
|e_x \rangle = \hat q |e_0 \rangle \;\,\mbox{with} \;\, \hat q = \hat Q\hat w^{-\alpha}/\langle e_0 |\hat Q\hat w^{-\alpha}|e_0 \rangle
\label{sud1a}
\nq
and $\langle e_x |=\langle e_0 |$. In the model (\ref{sud1}) one gets for any $\hat {\widetilde{W}}_i$
\bq
\hat {\widetilde{\cal R}}_i = \langle \hat {\widetilde{P}}_{Q_i} \rangle + \langle {\hat q^{-1}\widetilde{P}}_Q \rangle [1 - \langle \hat q^{-1} \hat {\widetilde{W}}_Q \rangle]^{-1} \langle \hat {\widetilde{W}}_{Q_i} \rangle,
\label{sad3}
\nq
where $\hat {\widetilde{P}}_{Q_i} = (1 - \hat {\widetilde{W}}_{Q_i})/\hat \Omega,\,
\hat {\widetilde{W}}_Q = (1+ \hat \Phi \hat Q)^{-1},\,$ and $\,\hat {\widetilde{W}}_{Q_i} = \hat {\widetilde{W}}_i (\hat {\widetilde{W}}_Q/\hat {\widetilde{W}})$.

{\it Short correlation time limit (SCTL).}\,
In practical applications of special importance is the SCTL for $V (t)$-fluctuations in which eq. (\ref{sad3}) can be markedly simplified. It corresponds to large characteristic rates $w_c $ of the dependence $\hat \Phi (\hat \Omega) \equiv \hat \Phi (\hat \Omega/w_c)$: $w_c \gg \|  V \|$. In this limit the relaxation kinetics is described by the first terms of the expansion of $\hat \Phi (\hat \Omega/w_c)$ in small $ \hat \Omega /w_c $, since $\hat \Phi (\epsilon)$ is the increasing function of $\epsilon$ with $\hat \Phi (\epsilon) \stackrel{\epsilon \to 0}{\longrightarrow} 0$ . Some important general conclusions, however, can be made independently of the form of $\hat \Phi (\Omega)$ (see below).

{\it Models for quantum evolution and fluctuations.}\,
The obtained general results are conveniently illustrated with the quantum two-level model and the stochastic two-state SRM for $V(t)$-fluctuations.

Quantum evolution of the two-level system is governed by the hamiltonian (assumed to be real matrix)
\bq
H_s =
\frac{\omega_s}{2} \left[%
\begin{array}{rr}
  1 & 0 \\
  0 & -1 \\
\end{array}%
\right], \;
{\cal V} =
\left[%
\begin{array}{cc}
  {\cal V}_d & {\cal V}_n  \\
  {\cal V}_n  & -{\cal V}_d \\
\end{array}%
\right] \: %
\begin{array}{c}
  |+\rangle \\
  |-\rangle \\
\end{array}.%
\label{two1}
\nq

The two-state SRM suggests that fluctuations result from jumps between two states (in $\{x \}$-space), say, $|x_+\rangle$ and $|x_-\rangle$ whose kinetics is described by
\bq
\hat {\cal L} = 2 (1 - |e_x \rangle \langle e_x |) \;\:\mbox{with} \;\: | e_x \rangle = \mbox{$\frac{1}{2}$}\big||x_+\rangle + |x_-\rangle\big\rangle.
\label{two2}
\nq

Below we will consider two examples of these models:

1) \underline{Diagonal noise} \cite{And}: $\omega_s = 0$, $\, {\cal V}_n = 0, \, {\cal V}_d =  \omega_0 (|x_+\rangle \langle x_+| - |x_-\rangle \langle x_-|)$, and
\bq
H_{\nu = \pm} = \pm \mbox{$\frac{1}{2}$}\omega_0 (|+\rangle \langle+|\, - \,|-\rangle \langle-|);
\label{two3}
\nq

2) \underline{Non-diagonal noise}: ${\cal V}_d = 0\:$ and $\: {\cal V}_n = v (|x_+\rangle \langle x_+| - |x_-\rangle \langle x_-|)$, so that
\bq
H_{\nu = \pm} = H_s \,\pm\, v (|+\rangle \langle-|\, + \,|-\rangle \langle+|).
\label{two4}
\nq

The first model describes dephasing while the second is useful for studying population relaxation.

In the model (\ref{two1}) dephasing and population relaxation are characterized by two functions: the spectrum $I (\omega)$ 
and the difference of level populations $N(t)$:
\bq
I (\omega) = \mbox{$\frac{1}{\pi}$}{\rm Re}\langle s | \hat {\widetilde{\cal R}} (i\omega) |s\rangle
\;\:\mbox{and} \;\: N(t) = \langle n | \hat {{\cal R}} (t) |n\rangle
\label{two5}
\nq
with
$|s \rangle = \frac{1}{\sqrt{2}}\big||\!+\!- \rangle + |\!-\!+ \rangle \! \big\rangle$ and  $ |n \rangle = \frac{1}{\sqrt{2}}\big||\!+\!+ \rangle - |\!-\!- \rangle \!\big\rangle$.

\bigskip

{\bf General results in the SCTL.}\,
Within the SCTL ($\|V \|/w_c \ll 1$) especially simple results are obtained for $\|H_s\|/ w_c \ll 1$. In the lowest order in $\|\hat \Phi (\hat \Omega/w_c)\| \ll 1$
\bq
\hat {\widetilde{\cal R}} \approx \hat {\widetilde{\cal R}}_n \approx \langle \hat q^{-1}\hat Q {\hat \Omega}^{\!-1} \hat \Phi (\hat \Omega)\rangle / \langle \hat q^{-1}\hat Q \hat \Phi (\hat \Omega) \rangle.
\label{sctl1}
\nq
This formula holds for any initial matrix $\hat {\widetilde{W}}_i$ and, in particular, for $s$-fluctuations, if $\|\hat t_w \| \sim 1/w_c \ll  1/\|\hat \Omega \|$.

The more complicated SCTL-case $ \| H_s \|/w_c \simeq 1$ can be analyzed by expanding $\hat {\widetilde {\cal G}}$ in powers of the parameter $\xi = \| V \|/\| H_s \| \ll 1$. In particular, within the general two-level model [eq. (\ref{two1})] with $V_d = 0$ in the second order in $\xi$ the diagonal and non-diagonal elements of $\rho (t)$ are decoupled and the corresponding elements of $\hat {\cal R} (t)$ are expressed in terms of the universal function
\begin{eqnarray}
R_{k} (t) &=& \frac{1}{2\pi i}\int_{-i\infty}^{i\infty}\!d\epsilon \,\frac{e^{i\epsilon t}}{\epsilon + k \epsilon/\langle \hat \Phi (\epsilon)\rangle}:
\label{sctl4a}\\
\langle \mu |\hat {\cal R}(t)|\mu \rangle &=& e^{-i\omega_{\mu} t} R_{k_{\mu}} (t), \;\;(\mu =n, +-,\, -+ ), \label{sctl4b}
\end{eqnarray}
where $\omega_{\mu} = \langle \mu |\hat { H}_{s}|\mu \rangle,\, k_n = 2{\rm Re} (k_{+-}),\,$ and
\bq
k_{+-} = k_{-+}^{*} = \mbox{$\frac{1}{2}$}\omega_s^{-2} \langle {\cal V}_n\hat q^{-1}[1-\hat {\widetilde W}_Q (2i\omega_s)] {\cal V}_n \rangle.
\label{sctl2}
\nq

\medskip
{\bf Anomalous fluctuations.}
The simplest model for anomalous fluctuations can be written as \cite{Met}
\bq
\hat \Phi (\epsilon) = (\epsilon/\hat w)^{\alpha}, \; (0 < \alpha < 1),
\label{res1}
\nq
where $\hat w$ is the matrix of fluctuation rates diagonal in $|x\rangle$-basis. For simplicity, $\hat w$ is assumed to be independent of $x$, i.e. $\hat w \equiv w (= w_c)$. The model (\ref{res1}) describes anomalously slow decay of the matrix $\hat W (t) \sim 1/t^{1+\alpha}$ (very long memory effects in the system \cite{Met}), for which only the case of $n$-fluctuations is physically sensible.

In the SCTL (\ref{sctl1}) the model (\ref{res1}) yields the expression
\bq
\hat {\widetilde{\cal R}}_n (\epsilon) = \langle \hat \Omega^{\alpha - 1}(\epsilon) \rangle  \langle \hat \Omega^{\alpha}(\epsilon) \rangle^{-1} \; \mbox{with} \;\, \hat \Omega (\epsilon) = \epsilon + i\hat {\cal H}
\label{res2}
\nq
which shows that  $\hat {\widetilde{\cal R}}_n (\epsilon)$ [and $\hat {{\cal R}}_n (t)$] is independent of the characteristic rate $w$. For $\alpha = 0$ and $\alpha = 1$ eq. (\ref{res2}) reproduces the static and fluctuation narrowing limits \cite{Abr}: $\hat {\widetilde{\cal R}}_n (\epsilon) = \langle \hat \Omega^{- 1} (\epsilon) \rangle$ and $\hat {\widetilde{\cal R}}_n (\epsilon) = 1/\langle \hat \Omega (\epsilon) \rangle$, respectively.

Of certain interest is the limit $\alpha \to 1$ in which formula (\ref{res2}) predicts the Bloch-type exponential relaxation:
\bq
\hat {\widetilde{\cal R}}_n (\epsilon) \approx [\epsilon + i \hat {H}_s +
(\alpha -1)\langle \hat {\Omega} \ln (\hat {\Omega})\rangle_{\epsilon \to 0}
]^{-1},
\label{res2a}
\nq
controlled by the relaxation rate matrix $\hat W_r = (\alpha-1){\rm Re}\langle \hat {\Omega} \ln (\hat {\Omega})\rangle_{\epsilon \to 0}$, and accompanied by frequency shifts represented by $\hat h = i(\alpha -1){\rm Im}\langle \hat {\Omega} \ln (\hat {\Omega})\rangle_{\epsilon\to 0}$. However, the matrices $\hat W_r$ and $\hat h$ (unlike those in the conventional Bloch equation) are independent of the characteristic rate $w$ of $V(t)$-fluctuations.

\paragraph{Dephasing for diagonal noise.}
In the model (\ref{two3}) the spectrum $I(\omega)$ can be obtained in the general SRM (\ref{sud1}):
\bq
I(\omega) = n_{\alpha} \frac{ \psi_{-}^{\alpha} \psi_{+}^{\alpha-1} + \psi_{-}^{\alpha-1}
\psi_{+}^{\alpha}}{(\psi_{-}^{\alpha})^2 + (\psi_{+}^{\alpha})^2 + 2  \psi_{-}^{\alpha}
\psi_{+}^{\alpha} \cos (\pi \alpha)},
\label{res3}
\nq
where $\,\psi_{\pm}^{\beta} (\omega) = \langle |\omega - 2V_d |^{\beta} \theta [\pm (\omega - 2V_d)]\rangle \, $ with $\theta (z)$ being the Heaviside step-function and $n_{\alpha} = \sin (\pi \alpha)/\pi$.

In the two-state SRM (\ref{two3})
$I(\omega) = \frac{1}{2}n_{\alpha}\omega_0^{-1} \theta(y)\,
(y+y^{-1}+2)/[y^{\alpha}+y^{-\alpha}+2\cos(\pi\alpha)],\, $
where $y = (\omega_0 + \omega)/(\omega_0 - \omega)$ (see also ref. \cite{Silb}). According to this formula anomalous dephasing (unlike conventional one \cite{Abr}) leads to broadening of $I(\omega)$ only in the region $|\omega| < \omega_0$ and singular behavior of $I (\omega)$ at $\omega \to \pm \omega_0$: $I(\omega) \sim 1/(\omega \pm \omega_0)^{1-\alpha}$. For $\alpha > \alpha_c \approx 0.59$ [$\alpha_c$ satisfies the relation $\alpha_c = \cos (\pi \alpha_c/2)$] the two-state-SRM formula also predicts the occurrence of the central peak (at $\omega = 0$) \cite{Silb} of Lorenzian shape and width $w_L \approx \omega_0 \cos (\pi \alpha/2)/\sqrt{\alpha^2-\cos^2(\pi \alpha/2)}$: $\: I (\omega) \approx \frac{1}{2\pi}\tan(\pi \alpha/2)\omega_0^{-1}/[1 + (\omega/w_L)^2]$, whose intensity increases with the increase of $\alpha - \alpha_c$ (Fig. 1). At $\alpha \sim 1$ the parameters of this peak are reproduced by eq. (\ref{res2a}) in which $\langle \hat {\Omega} \ln (\hat {\Omega})\rangle_{\epsilon} = -(\pi/2)\omega_0$. The origination of the peak indicates the transition from the static broadening at $\alpha \ll 1$ to the narrowing one at $\alpha \sim 1$ [see eq. (\ref{res2})]. For systems with complex spectra this transition can, of course, be strongly smoothed.

\paragraph{Dephasing for non-diagonal noise.}
The model (\ref{two4}) allows one to reveal some additional specific features of dephasing. We restrict ourselves to the analysis of the case $\|H_s\| \sim \omega_s \gtrsim w$ and the most interesting part of the spectrum at $|\omega | \sim \omega_s$. Equations (\ref{sctl4a}) and (\ref{sctl4b}) show that in this case the elements $\langle \mu |{\cal R} (t) | \mu \rangle ,\,(\mu = +-,\,-+),$ which describe phase relaxation are given by $\,  \langle \mu |{\cal R} (t) | \mu \rangle = e^{-i\omega_{\mu} t } E_{\alpha} [-k_{\mu}(w t)^{\alpha}],$ where  $E_{\alpha} (-z) = (2\pi i)^{-1}\int_{-i\infty}^{i\infty}\! dy\, e^{y}/(y + z y^{1-\alpha})$ is the Mittag-Leffler function \cite{Met}. Therefore for $|\omega | \sim \omega_s$
\bq
I (\omega) = I_0 (\omega_s + \omega) + I_0 (\omega_s - \omega),
\label{res5}
\nq
where
\bq
I_0 (\omega) = n_0 \sin \!\phi_x(|x|^{1+\alpha} + |x|^{1-\alpha} + 2|x|\cos \!\phi_x )^{\!-1}
\label{res5a}
\nq
with  $x = \omega/(|k_{+ -}|^{1/\alpha}w),\;$ $n_0 = (\pi |k_{+ -}|^{1/\alpha}w)^{-1},$ and
\bq
\phi_x = \frac{\pi\alpha}{2} +{\rm sign} (x) \arctan \!\biggl[\frac{\sin (\frac{1}{2}\pi\alpha)}{\cos(\frac{1}{2}\pi\alpha)+2^{-\alpha^{-1}}\omega_s/w}\!\biggr].\nonumber
\nq
Formula (\ref{res5}) predicts singular behavior of $I (\omega)$ at $\omega \sim \pm\omega_s$: $I(\omega) \sim 1/|\omega\pm\omega_s|^{1-\alpha}$, and slow decrease of $I (\omega)$ with the increase of $|\omega \pm \omega_s|$: $I(\omega) \sim 1/|\omega \pm \omega_s|^{1+\alpha}$.

In the limit $\omega_s/w \ll 1\:$ $\phi_x \approx \pi\alpha \theta (x)$ so that $I_0 (\omega) \sim \theta (\omega)$. This means that for $\omega_s/w \ll 1\:$ the spectrum $I (\omega)$ is localized in the region $|\omega | < \omega_s$ and looks similar to $I (\omega)$ for diagonal dephasing at $\alpha < \alpha_c$ (see Fig. 1). For $\omega_s/w \gtrsim 1$, however, $I (\omega)$ is non-zero outside this region, moreover, in the limit $\omega_s/w \gg 1$ the spectrum $I_0 (\omega)$ becomes symmetric: $I_0 (\omega) = I_0 (-\omega)$, similar to the conventional spectra.

It is also worth noting that for $\omega_s/w \ll 1\,$ functions $\langle \mu |{\cal R} (t) | \mu \rangle$ and $I(\omega)$ are independent of $w$  [in agreement with eq. (\ref{sctl1})] since $k_{\mu} \sim (\omega_s/w)^{\alpha}$ and $k_{\mu}(w t)^{\alpha} \sim (\omega_s t)^{\alpha}$. In the opposite limit, however, $k_{\mu} \sim w^{0}$ so that the characteristic relaxation time $\sim w^{-1}$.

\paragraph{Population relaxation.}
Specific features of anomalous population relaxation can be analyzed with the model of non-diagonal noise (\ref{two4}).

In particular, in the limits $\|H_s\| \sim \omega_s \gtrsim w$ and $1-\alpha \ll 1$ with the use of eqs. (\ref{sctl4a}),(\ref{sctl4b}) and (\ref{res2a}) one gets
\bq
N(t) = E_{\alpha} [-k_n(w t)^{\alpha}] \;\;\mbox{and} \;\; N(t) = e^{- w_{\alpha} t},
\label{res6}
\nq
respectively, where $ E_{\alpha} (-x)$ is the Mittag-Leffer function defined above and $w_{\alpha} \approx k_n (\alpha\to 1)w \sim 1-\alpha$. The first of these formulas predicts very slow population relaxation at $t > \tau_r = w^{-1}(k_n/w)^{1/\alpha}$: $N (t) \sim 1/t^{\alpha}$. Similar to $I(\omega)$ the function $N (t)$ is, in fact, independent of $w$ in the limit $\omega_s/w \ll 1$ because in this case $k_n \sim (\omega_s/w)^{\alpha}$. In the opposite limit $\omega_s/w > 1$ the characteristic time population relaxation is $\sim w^{-1}$ since $k_n$ is independent of $w$ as with the phase relaxation.

In the limit $\|H_s\|, \|V\| \ll w$ one obtains
\bq
N (t) = \frac{1}{2\pi i} \int_{-i\infty}^{i\infty}\!\! d\epsilon\, e^{\epsilon t}\,\frac{\omega_s^2 \epsilon^{\alpha-1} + 4v^2\Omega_{\alpha-1}(\epsilon)}{\omega_s^2 \epsilon^{\alpha} + 4v^2\Omega_{\alpha}(\epsilon)},
\label{res7}
\nq
where $\Omega_{\beta}(\epsilon) = [(\epsilon + 2i E_0)^{\beta}+(\epsilon - 2i E_0)^{\beta}]/2$ and $E_0 = \sqrt{v^2 + \omega_s^2/4}$. Naturally, in the corresponding limits expression (\ref{res7}) reproduces formulas (\ref{res6}) with $k_n \approx 2^{\alpha-1}\cos(\pi\alpha/2)(E_0/w)^{\alpha}$ and $w_{\alpha} \approx \pi (1-\alpha)v^2/E_0$ (Fig. 2). Outside these limits $N (t)$ can be evaluated numerically (some results are shown in Fig. 2). In general, $N (t)$ is the oscillating function (of frequency $\sim E_0$) with slowly decreasing average value and oscillation amplitude: for $E_0 t \gg 1$  $N (t) \sim 1/t^{\alpha}$ (except the limit $\alpha \to 1$).

\medskip
{\bf Concluding remarks.}\, The presented analysis of relaxation kinetics in quantum systems induced by anomalous noise demonstrates a number of peculiarities of this kinetics. The peculiarities are analyzed with the use of the two-level quantum model, as an example, though the observed anomalous effects can show themselves in more complicated multi-level quantum systems. The proposed theoretical method is quite suitable for the analysis of these systems. This work is currently in progress.

Noteworthy is that in some limits the theory developed predicts relaxation kinetics described by the Mittag-Leffler function $E_{\alpha} [-(wt)^{\alpha}]$. Following a number of recent works (for review see ref. \cite{Met}) this kinetics can be thought as a result of the anomalous Bloch equation with fractional derivative in time. For brevity we have not discussed the corresponding representations.

It is also interesting to note that with the increase of $\alpha$ the effects of anomaly of fluctuations decrease but still persist. To clarify them let us briefly consider the model
$\Phi (\epsilon) = (\epsilon /w) + \zeta (\epsilon /w)^{1+\alpha},$ in which $0 < \alpha < 1$, and $w$ and $\zeta$ are the constants with $\zeta \ll 1$ [small value of $\zeta$ ensures that $W (t) > 0$]. Possible effects can be analyzed within the SCTL with the use of eqs. (\ref{sctl1})-(\ref{sctl4b}). For example, in the limit $\|H\|/w \ll 1$ one obtains formula
$\,\widetilde{\cal R} \approx [\epsilon + i \hat H_s + \zeta w^{-\alpha} \langle (i\hat {\cal H})^{1+\alpha} - (i\hat H_s)^{1+\alpha}\rangle]^{-1}\,$
predicting the Bloch-type relaxation of both phase and population, but with the rate $\hat W_r = \zeta w^{-\alpha} {\rm Re}  \langle (i\hat {\cal H})^{1+\alpha} - (i\hat H_s)^{1+\alpha}\rangle $ which depends on $w$ as $w^{-\alpha}$, i.e. slower than in the conventional Bloch equation ($\hat W_r \sim 1/w$ \cite{Abr}). Analysis also shows that in the expression for $\widetilde{\cal R}$ the terms $\sim w (\epsilon/w)^{1+\alpha}$ occur as well. They lead to the inverse power-type asymptotic behavior of $\langle \mu | \hat R (t)|\mu \rangle \sim 1/t^{2+\alpha}$ observed, however, only at very long times $t \gg w^{-1}$.

In our brief analysis we neglected the effect of possible natural width of lines corresponding to the additional slow exponential relaxation in the system. It is clear that the developed method allows one take in account these effects straightforwardly in case of need.

\medskip
{\bf Acknowledgements.}\, The work was partially supported by the Russian Foundation for Basic Research.

{\bf Figures.}

Fig. 1.The spectrum $I (x) = I(\omega)\omega_0$, where $x = \omega/\omega_0$, calculated in the model (\ref{two3}) [using eq. (\ref{res3})] for different values of $\alpha$: (1) $\alpha = 0.5$, (2) $\alpha = 0.7$, (3) $\alpha = 0.8$, and (4) $\alpha = 0.9$.


Fig. 2. Population relaxation kinetics $N (\tau)$, where $\tau = E_0 t$, calculated with eq. (\ref{res7}) (a) for large $\alpha$ and different $r = 2v/\omega_s$: (1)-$\alpha = 0.95, \: r = 1.0;\:$ (2)-$\alpha = 0.95, \: r = 2.0;\:$ (3)-$\alpha = 0.88, \: r = 1.0;\:$ $\alpha = 0.88, \: r = 2.0;\:$ and (b) for small $\alpha$ ($r = 0.7$): $\alpha = 0.3$ (full lines) and $\alpha = 0.5$ (dotted lines). Straight lines in figures (a) and (b) represent exponential [eq. (\ref{res6})] and $t^{-\alpha}$ dependencies, respectively (in Fig. 2a they are shown by dashed lines).





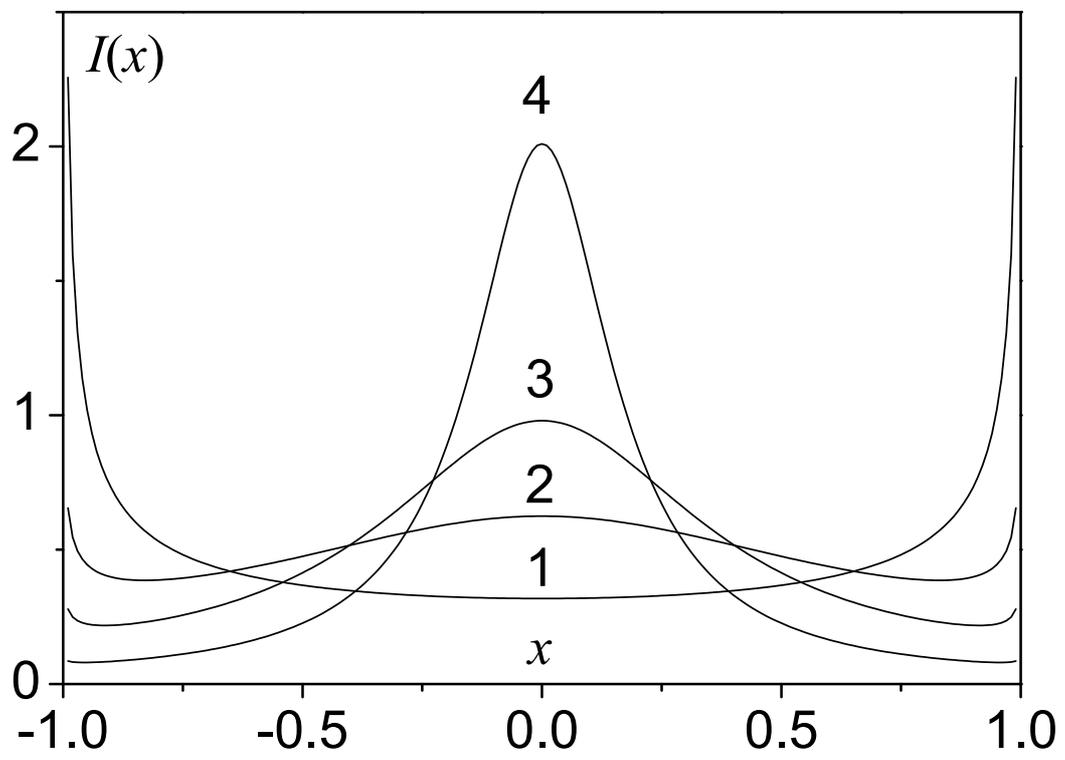
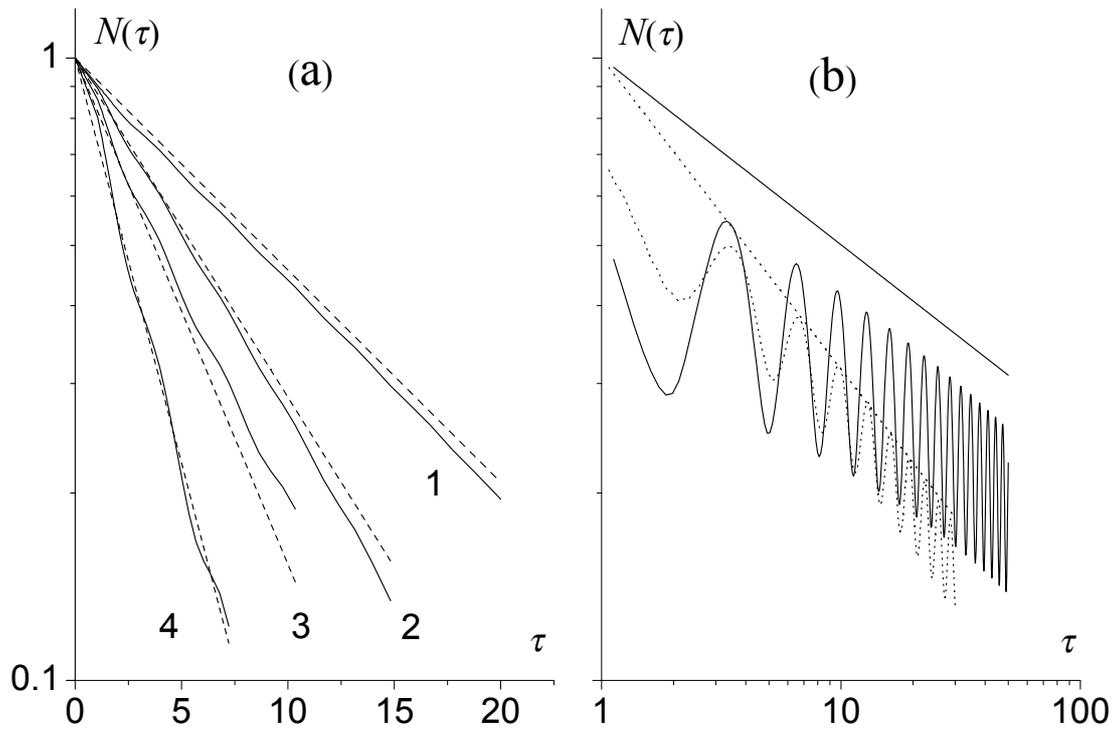
\end{document}